\documentclass{pasj00}
\draft

\begin{document}
\SetRunningHead{Aya Bamba et al.}{SN~1006 with {\it Suzaku}}
\Received{}
\Accepted{}

\title{Suzaku wide-band observations of SN~1006}

\author{Aya \textsc{Bamba}\altaffilmark{1},
Yasushi \textsc{Fukazawa}\altaffilmark{2},
Junko S. \textsc{Hiraga}\altaffilmark{3},
John P. \textsc{Hughes}\altaffilmark{4},
Hideaki \textsc{Katagiri}\altaffilmark{2},
Motohide \textsc{Kokubun}\altaffilmark{1},
Katsuji \textsc{Koyama}\altaffilmark{5},
Emi \textsc{Miyata}\altaffilmark{6},
Tsunefumi \textsc{Mizuno}\altaffilmark{2},
Koji \textsc{Mori}\altaffilmark{7}
Hiroshi \textsc{Nakajima}\altaffilmark{6}
Masanobu \textsc{Ozaki}\altaffilmark{1},
Rob \textsc{Petre}\altaffilmark{8},
Hiromitsu \textsc{Takahashi}\altaffilmark{2},
Tadayuki \textsc{Takahashi}\altaffilmark{1},
Takaaki \textsc{Tanaka}\altaffilmark{9},
Yukikatsu \textsc{Terada}\altaffilmark{3},
Yasunobu \textsc{Uchiyama}\altaffilmark{1},
Shin \textsc{Watanabe}\altaffilmark{1},
Hiroya \textsc{Yamaguchi}\altaffilmark{5}
}
\altaffiltext{1}{Department of High Energy Astrophysics,
Institute of Space and Astronautical Science (ISAS),\\
Japan Aerospace Exploration Agency (JAXA),\\
3-1-1 Yoshinodai, Sagamihara, Kanagawa 229-8510, Japan}
\altaffiltext{2}{Department of Physics, Hiroshima University
Higashi-Hiroshima, 739-8526, Japan}
\altaffiltext{3}{RIKEN, cosmic radiation group,
2-1, Hirosawa, Wako-shi, Saitama, Japan}
\altaffiltext{4}{Department of Physics and Astronomy, Rutgers University,\\
136 Frelinghuysen Road, Piscataway, NJ 08854-8019, USA}
\altaffiltext{5}{Department of Physics, Graduate School of Science,
Kyoto University, \\
Sakyo-ku, Kyoto 606-8502, Japan}
\altaffiltext{6}{Department of Astrophysics, Faculty of Science,
Osaka University, Toyonaka 560-0043, Japan}
\altaffiltext{7}{Department of Applied Physics, Faculty of Engineering
University of Miyazaki,\\
1-1 Gakuen Kibana-dai Nishi, Miyazaki, 889-2192, Japan}
\altaffiltext{8}{NASA Goddard Space Flight Center, Greenbelt, MD 20771, USA}
\altaffiltext{9}{Stanford Linear Accelerator Center,
2575 Sand Hill Road M/S 29 Menlo Park, CA 94025 USA}
\email{(AB): bamba@astro.isas.jaxa.jp}


%


\maketitle

\begin{abstract}
We report on the wide band spectra of SN~1006 as observed by Suzaku.
Thermal and nonthermal emission are successfully resolved
thanks to the excellent spectral response of Suzaku's X-ray CCD XIS.
The nonthermal emission cannot be reproduced
by a simple power-law model
but needs a roll-off at 5.7$\times 10^{16}$~Hz
= 0.23~keV.
The roll-off frequency is significantly higher in the northeastern rim
than in the southwestern rim.
We also have placed the most stringent upper limit
of the flux above 10~keV
using the Hard X-ray Detector.
\end{abstract}

\section{Introduction}
\label{sec:intro}

Since the discovery of cosmic rays nearly 100 years ago, their
origin and acceleration mechanism has remained largely a mystery.
The most widely accepted acceleration mechanism,
at least for cosmic rays up to $\sim 10^{15.5}$~eV (the knee energy),
is diffusive shock acceleration at the rims of young supernova
remnants (SNRs)
(e.g., \cite{bell1978, blandford1978}).
This hypothesis has received indirect observation confirmation from
the discovery using ASCA of synchrotron X-rays
from the shell of the SN~1006,  
implying the presence of electrons accelerated to relativistic energy
within the forward shocks  of SNRs \citep{koyama1995}.
Recently, \citet{bamba2003} suggested that
the acceleration efficiency of electrons is quite high
within the thin, filament-like shocks of SN~1006.
This appears to be a common feature in young SNRs
(e.g., \cite{vink2003}, \cite{yamazaki2004},
\cite{bamba2005}, \cite{ballet2005}, \cite{vink2006}).

One key observational feature in the spectrum of synchrotron emission is 
a roll-off energy,
which is determined by the maximum energy of accelerated electrons 
and the magnetic field.
There are several measurements of the roll-off energy
of the SN~1006 spectrum.
\citet{dyer2001} fit
combined Rossi X-Ray Timing Explorer (RXTE) and ASCA spectra
below 10~keV and measured a roll-off energy of
$3.0_{-0.2}^{+0.1} \times 10^{17}$~Hz.
The spatial dependence of the roll-off energy is reported
for the north-eastern rim \citep{bamba2003,allen2004}
and whole remnant \citep{rothenflug2004},
but the conditions producing the variation is unclear.
All these observations suffer from contamination
by the low energy thermal emission.
The 20--40~keV flux upper limit from the INTEGRAL ISGRI 
\citep{kalemci2006}
implies that the roll-off is lower than 4~keV.
To better measure the roll-off energy and its variation across the remnant,
it is crucial to make X-ray observations
with excellent statistics and energy response for diffuse emission,
in order to subtract thermal emission parameters properly.
The other need is the excellent sensitivity above 10~keV,
in order to measure the synchrotron spectrum over
as wide a band as possible.
Suzaku \citep{mitsuda2007} is the best observatory for such studies.
Suzaku has two sets of active instruments:
four X-ray Imaging Spectrometers (XIS, \cite{koyama2007})
each at the focus of an X-Ray Telescope (XRTs, \cite{serlemitsos2007})
and a separate Hard X-ray Detector (HXD, \cite{takahashi2007,kokubun2007}).
The XIS has two types of X-ray CCD,
3 front-illuminated (FI) CCDs and one back-illuminated (BI) one.
The former have high quantum efficiency (QE) in the 2--12~keV band,
whereas the BI CCD has high QE below 2~keV.
The HXD's silicon PIN diode array (hereafter PIN) has
a narrow field of view (30 arcmin on a side)
and an active background shield,
as a result, is the most sensitive detector
in the 10--70 keV band.

In this paper, we report the first results of
wide-band Suzaku observations of SN~1006.
\S\ref{sec:obs} introduces the observations, our analysis and results
is explained in \S\ref{sec:analysis},
and we discuss our findings in \S\ref{sec:discuss}.
We assume the distance of SN 1006 to be 2.17~kpc \citep{winkler2003}
in this paper.

\section{Observations and Data Reduction}
\label{sec:obs}

Suzaku observed SN~1006 and its background
with 8 pointings,
five for the SNR and three for the background,
during the performance verification (PV) phase.
We refer to each observation following the observation position,
as shown in table~\ref{tab:obslog}.
The data reduction and analysis were made
using HEADAS software version 6.1.2.,
version 1.2 of the processed data,
and XSPEC version 11.3.2.

The XIS was operated in the normal full-frame clocking mode
without spaced-row charge injection
in the all pointings.
We filtered out data obtained during passages through the
South Atlantic Anomaly (SAA),
with elevation angle to the Earth's limb below $\timeform{5D}$,
or with elevation angle to the bright Earth's limb below $\timeform{13D}$,
in order to avoid the contamination of emission from the bright Earth.
The XIS light curves in the NE observation show
an increased background rate 
during 2006 September 9th 19:16--20:23.
An X6.2 class solar flare was detected by the GOES satellite
during 2006 September 9th, 19:13--20:36\footnote{%
see http://www.sec.noaa.gov/}.
Thus, we ascribed the increased background rate to the solar flare
and filtered out the flaring time.
The net exposure times are listed in table~\ref{tab:obslog}.

The HXD PIN was operated in the normal mode.
We filtered out the data
obtained during passages through the
South Atlantic Anomaly,
with elevation angle to the Earth's limb below $\timeform{5D}$,
and cut off rigidity (COR) smaller than 8~GV.
The solar flare time in the NE observation is also filtered out.
The net exposure in each observation is in table~\ref{tab:obslog}.
The HXD team supplies two non-X-ray background (NXB) models,
{\tt METHOD = PINUDLC} (``bgd-a'')
and {\tt METHOD = LCFIT} (``bgd-d'')
(Mizuno et al.\ Suzaku-Memo 2006-42, Watanabe et al.\ Suzaku-Memo 2007-01,
Fukazawa et al.\ Suzaku-Memo 2007-02).
In both models, {\tt METHODV=1.2} version is adopted.
In this paper, we adopt the bgd-d model
as we explain in \S A\ref{sec:appendix}.

\section{Analysis}
\label{sec:analysis}

\subsection{XIS spectra}
\label{sec:xis-spec}

\subsubsection{Whole remnant}

\begin{figure}
\begin{center}
\FigureFile(140mm,50mm){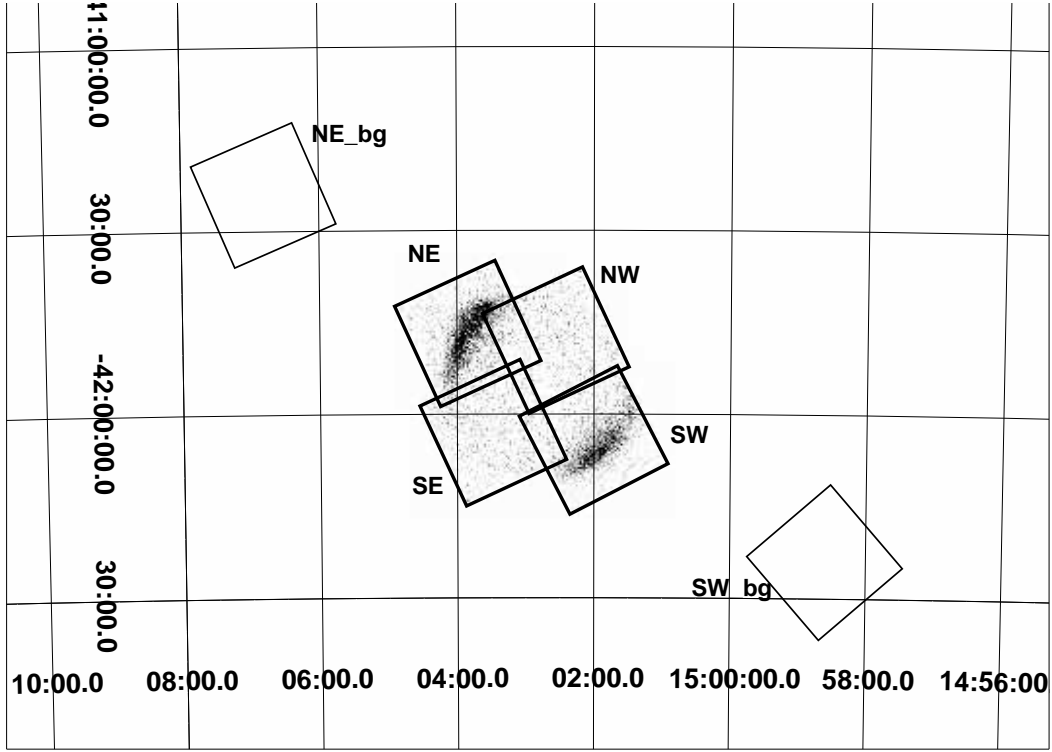}
\caption{XIS 2--7~keV image of SN 1006 (gray scale).
Boxes represent XIS fields of view for the various pointings.
Thick and thin box walls indicate rim and background observations, respectively.
SW1 and SW2, and SW\_bg1 and SW\_bg2 have almost same field of view
(see table~\ref{tab:obslog}),
and thus are marked just as SW and SW\_bg in this figure.}
\label{fig:regions}
\end{center}
\end{figure}

Figure~\ref{fig:regions} is the XIS 2.0--7.0~keV image,
which shows the nonthermal rims
on north-east (NE) and south-west (SW) regions.

\begin{figure}
\begin{center}
\FigureFile(80mm,40mm){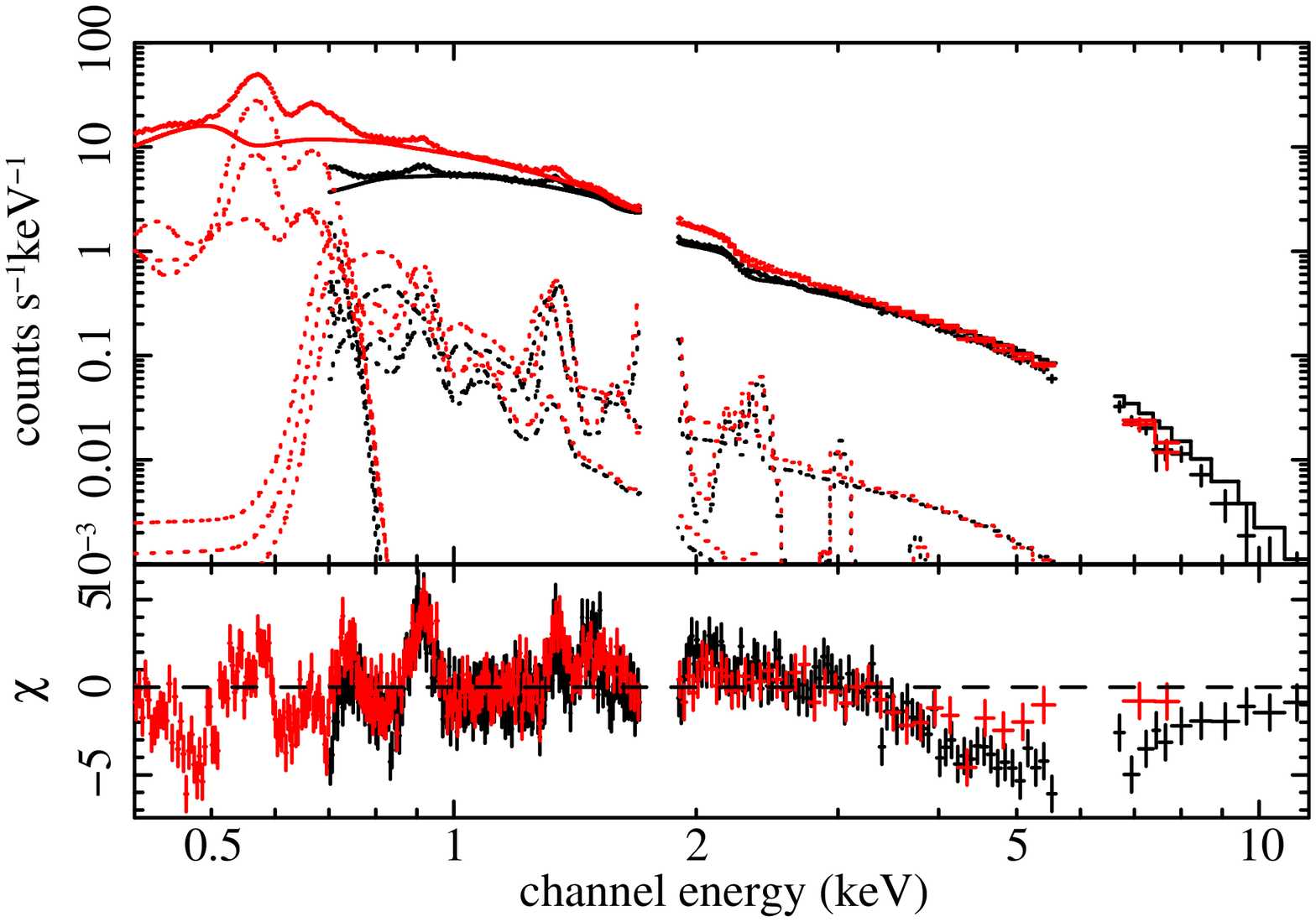}
\FigureFile(80mm,40mm){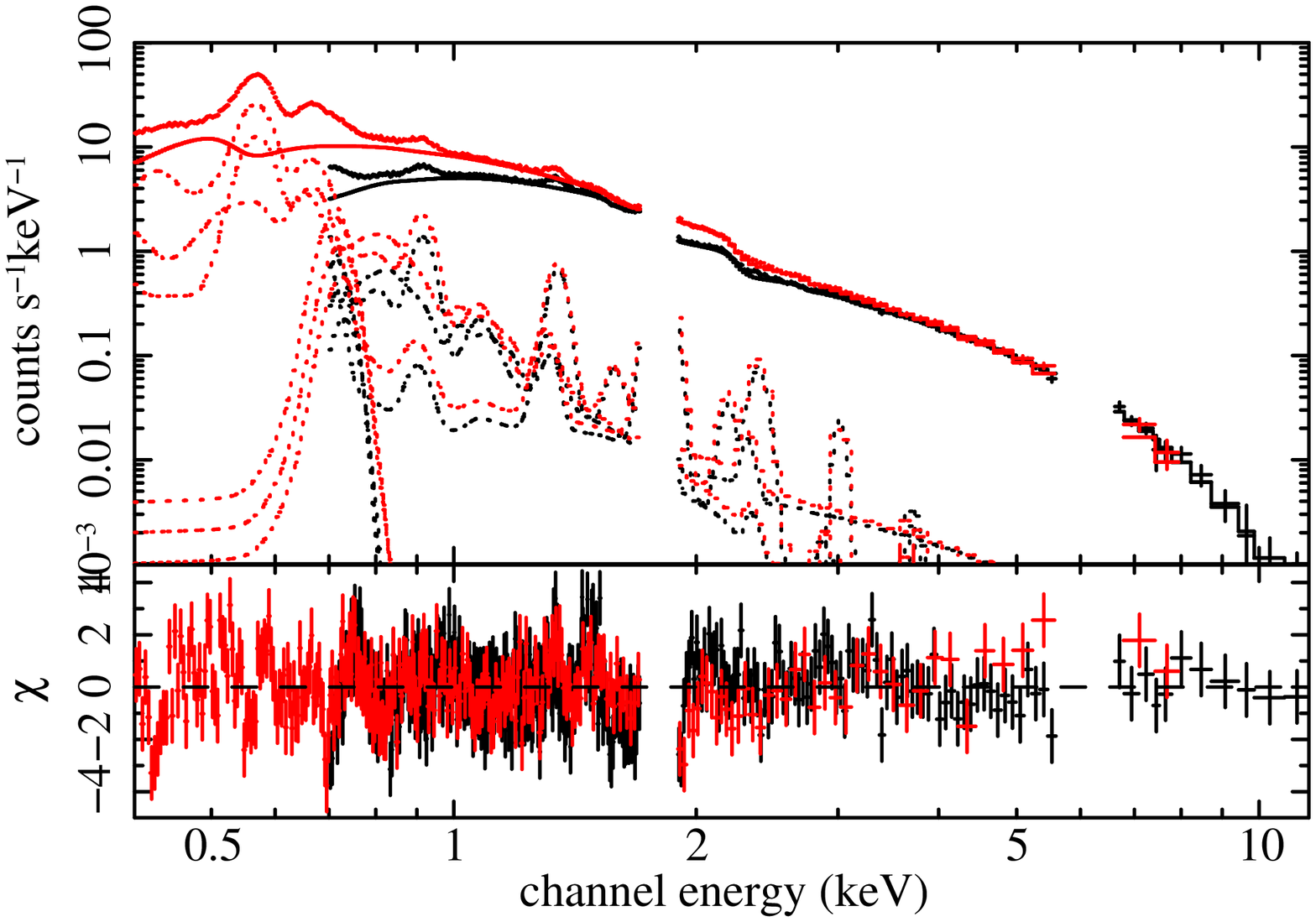}
\caption{XIS spectra with best-fit models of thermal + power-law (left)
or thermal + {\it srcut} (right).
Thermal and nonthermal models are represented with dotted or solid lines.
Black and red represents FI and BI spectra, respectively.
Lower panels in the figures are residuals from the best-fit models.}
\label{fig:xis_spec}
\end{center}
\end{figure}

We extracted combined spectra of the entire NE, SW1, and SW2 regions
(see table \ref{tab:obslog}),
since the nonthermal rims are in these fields of view.
The background spectrum was accumulated from the 3 background exposures
(NE\_bg, SW\_bg1, SW\_bg2).
We used 0.7--12.0~keV for the FI spectrum and 0.4--8.0~keV
for the BI one.
The 1.73--1.89~keV band is ignored, because the detector response 
function does not yet accurately model the Si edge.
We also ignored the 5.7--6.6~keV band
due to the contamination from internal calibration sources.
Figure~\ref{fig:xis_spec} shows the background-subtracted spectra.
In addition to strong emission lines below 2~keV,
a hard component is clearly detected up to 12~keV.
Therefore, we fitted the spectra using 
thermal plasma components plus a power-law continuum.
The response files were constructed using
{\tt xisrmfgen} in the HEADAS package.
The auxiliary response files were constructed using
0.5--10.0~keV XIS images and {\tt xissimarfgen} \citep{ishisaki2007}
in the HEADAS software package.
The thermal plasma model derived with Suzaku \citep{yamaguchi2007}
is used,
which has 3 non-equilibrium ionization components
(two ostensibly for ejecta and one for heated interstellar medium (ISM)).
For the absorption model,
we used the cross sections of
\citet{morrison1983} and assumed solar abundances \citep{anders1989}.
The absorption column was fixed at
$6.8\times 10^{20}$~cm$^{-2}$ following the H~I observation
 by \citet{dubner2002}.
The XIS team reported the gain uncertainty of $\sim \pm 5$~eV
(c.f.\ \cite{koyama2007}),
thus the gain is allowed to vary within a few percent separately
for the FI and BI spectra.

The fit to this initial model is statistically unacceptable,
with $\chi^2$/d.o.f. of 2513/589.
The residuals show line-like structure
in the low energy band,
so we tuned the temperature and the ionization parameter
of the heated interstellar medium components
and the S abundance of high temperature ejecta,
and fixed them again.
We also adjusted the normalization factor between FI and BI spectra
within a few percent.
\citet{serlemitsos2007} reported that
the systematic error of the telescope vignetting is $\sim$ 5\%,
so the few percent adjustment of the relative normalization is reasonable.
The best-fit models and parameters are shown
in figure~\ref{fig:xis_spec}(left) and table~\ref{tab:xis_para}.
The reduced $\chi^2$ improved to 2200/588,
but is still statistically unacceptable.
The residuals around the strong O lines are probably due to
the calibration uncertainty in the low energy band
\citep{koyama2007}.
The other large-scale residuals in the 2--10~keV band,
on the other hand, cannot be due to calibration uncertainties.
This result indicates that
the nonthermal component cannot be represented
by a simple power-law model.

We accordingly introduced the {\it srcut} model \citep{reynolds1998}
for the nonthermal component.
The {\it srcut} model represents
the synchrotron spectrum from an exponentially cut off power-law
distribution of electrons in a homogeneous magnetic field.
The radio spectral index is fixed to be $\alpha=0.57$
\citep{allen2001}.
The fit is still formally rejected
(see the right panel of figure~\ref{fig:xis_spec}
and table~\ref{tab:xis_para}),
but the reduced $\chi^2$ is substantially lower.
(857/588).
There are no large-scale residuals in the entire band.

\subsubsection{NE and SW rims}

\begin{figure}
\begin{center}
\FigureFile(80mm,40mm){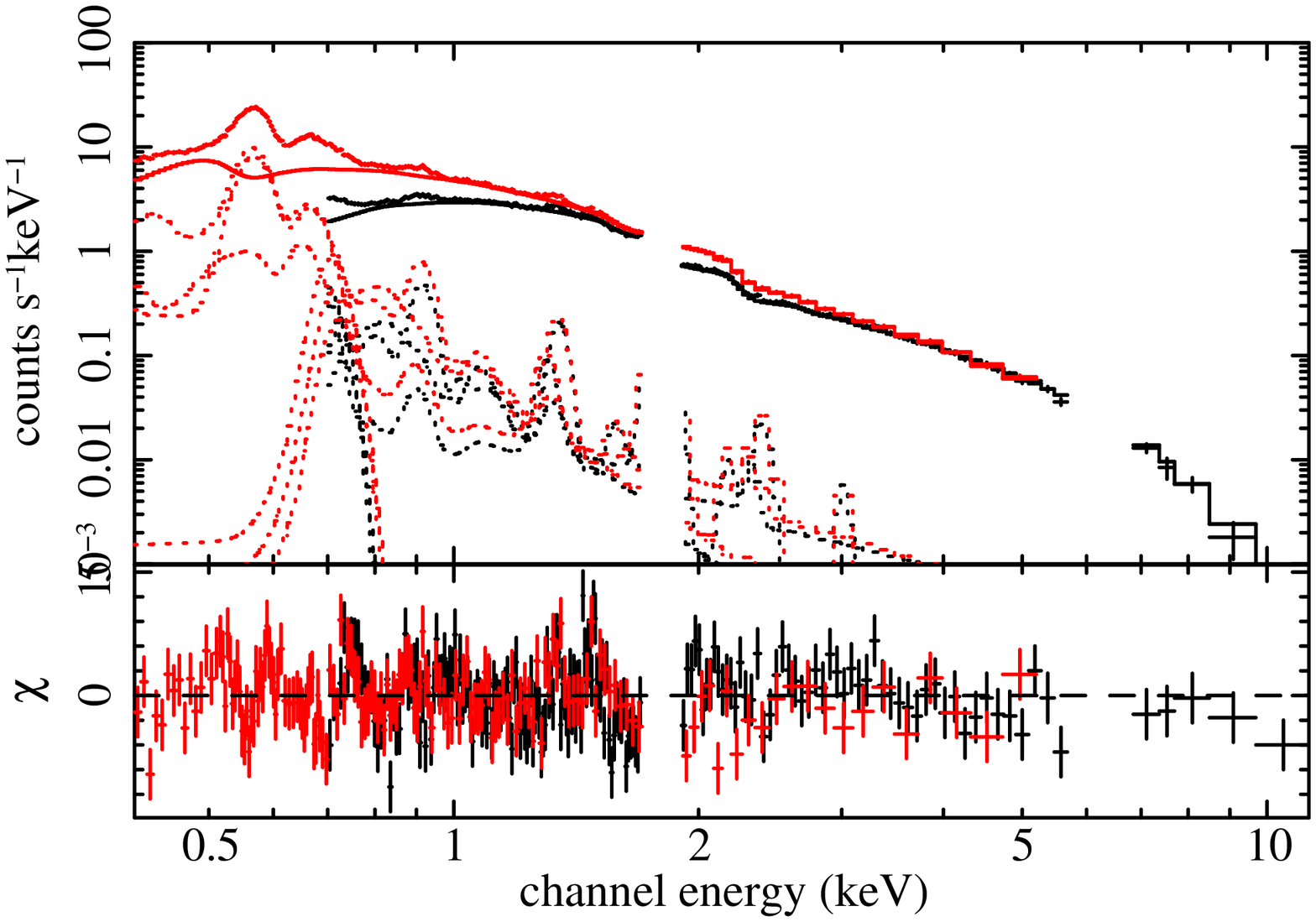}
\FigureFile(80mm,40mm){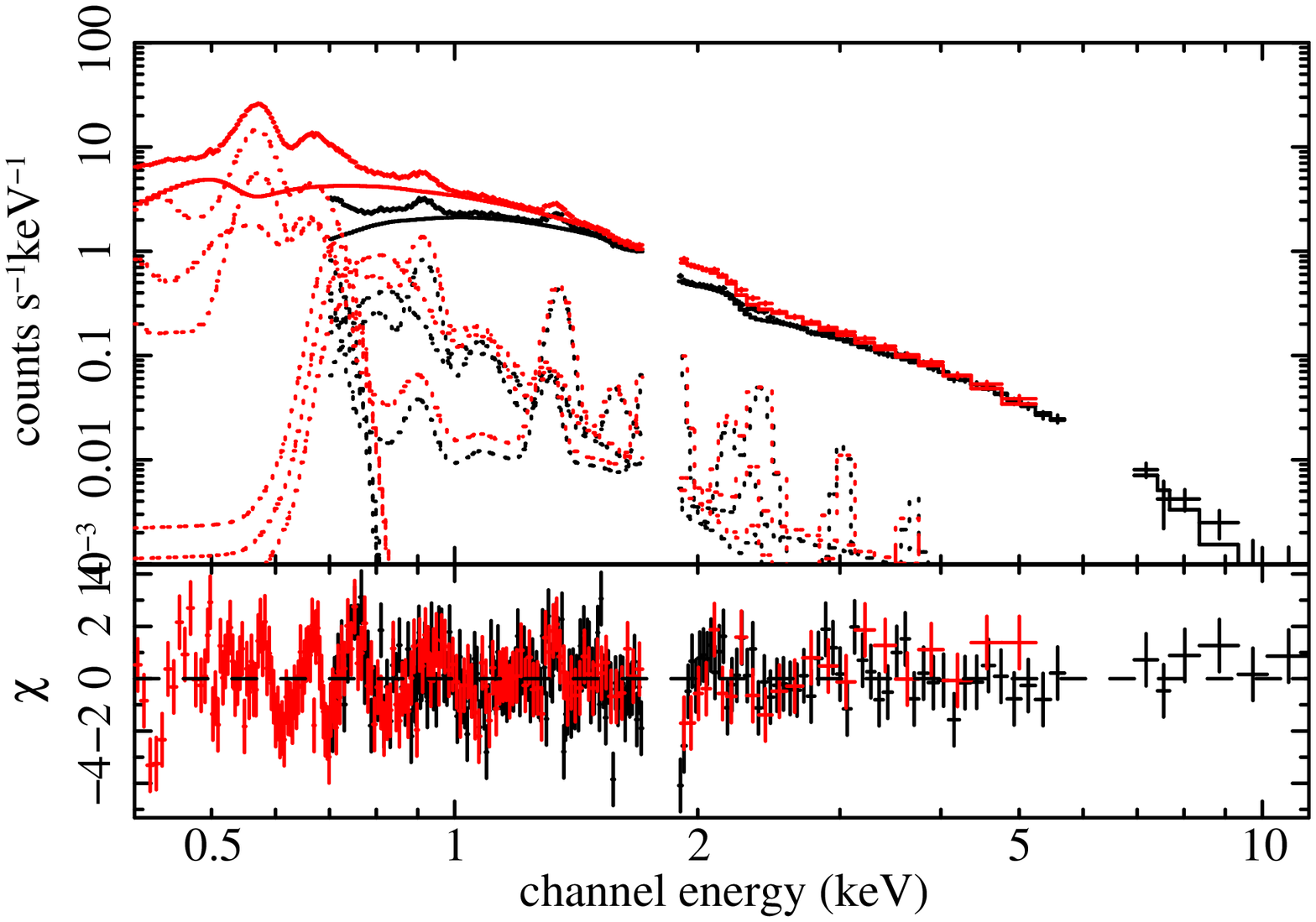}
\caption{XIS spectra of NE (left) and SW (right) regions.
Thermal and nonthermal models are represented with dotted or solid lines.
Black and red represents FI and BI spectra, respectively.
Lower panels in the figures are residuals from the best-fit models.}
\label{fig:xis_spec_parts}
\end{center}
\end{figure}

We applied the thermal model plus the {\it srcut} continuum
described above
to the NE and SW rim spectra separately.
The plasma parameters of the three thermal components were fixed
at the values for the whole remnant (see table~\ref{tab:xis_para})
except for the normalization.
The best-fit models and parameters are shown in
figure~\ref{fig:xis_spec_parts} and table~\ref{tab:xis_para_parts},
respectively.
The fittings are statistically unacceptable again
($\chi^2$ = 553/338 for NE and 527/368 for SW),
but they show no large-scale structure.

\subsection{HXD PIN spectra}
\label{sec:hxd-spec}

\begin{figure}
\begin{center}
\FigureFile(80mm,50mm){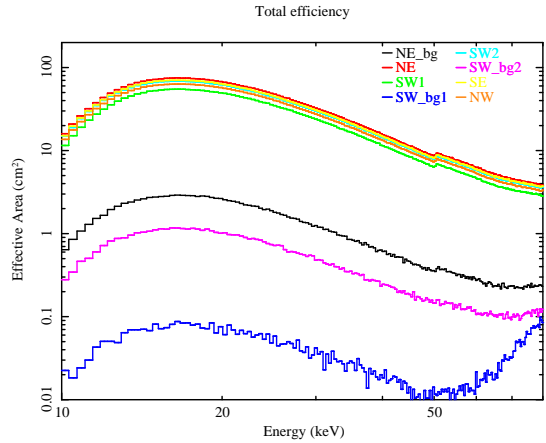}
\caption{The PIN effective area of each observation for the entire remnant.}
\label{fig:pin_rsp_hikaku}
\end{center}
\end{figure}

SN~1006 is an extended source for the PIN
as shown in figure \ref{fig:regions}.
We therefore have to consider the effect of the PIN angular response
for diffuse sources.
In order to estimate the total efficiency for the entire SNR,
we assumed that the emission region in the PIN energy band
is the same as that of ASCA GIS 2--7~keV image 
available from Data Archives and Transmission System
(DARTS\footnote{%
see http://darts.isas.jaxa.jp/astro/ .}),
which covers the entire remnant.
The derived efficiency in each observation is shown
in figure~\ref{fig:pin_rsp_hikaku}.
The discontinuities around $\sim$50~keV are due to the Gd K-line
back-scattered in GSO,
and the enhancement above 50~keV in the SW\_bg1 effective area
is due to the transparency of the passive shield which becomes
larger in the higher energy band
\citep{takahashi2007}.
\citet{takahashi2007b} checked
the influence of the source size on the effective area
for extended sources,
and found that it has no energy dependence.
The rim observations,
NE, SW1, SW2, SE, and NW,
have similar efficiency,
while the background observations 
(NE\_bg, SW\_bg1, and SW\_bg2) have
those of more than 1 order smaller.
Thus, we combined 5 rim observations
to improve statistics.
The 3 background observations are also combined
and used 
to evaluate the contribution of the cosmic X-ray background (CXB).

\begin{figure}
\begin{center}
\FigureFile(80mm,40mm){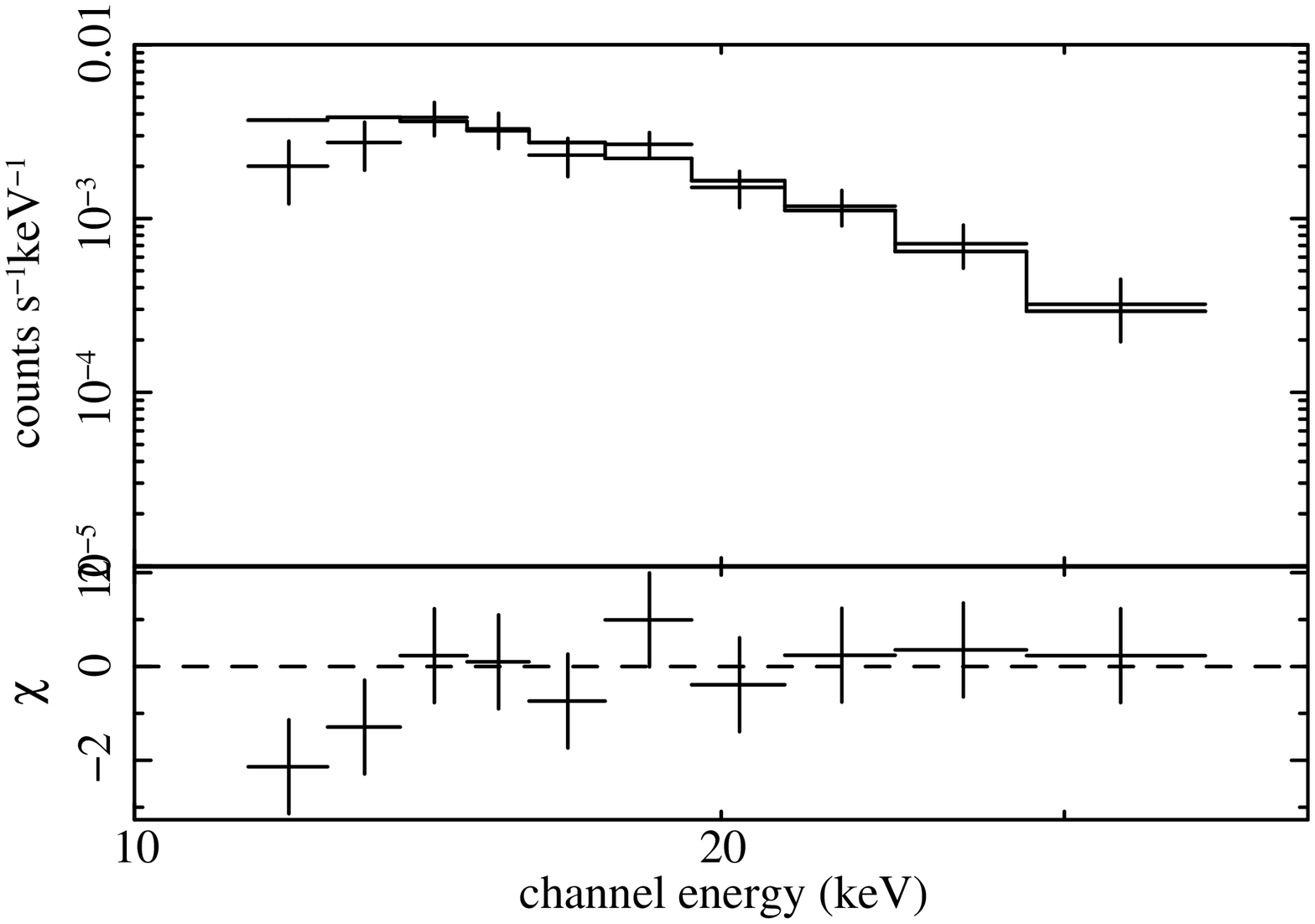}
\FigureFile(80mm,40mm){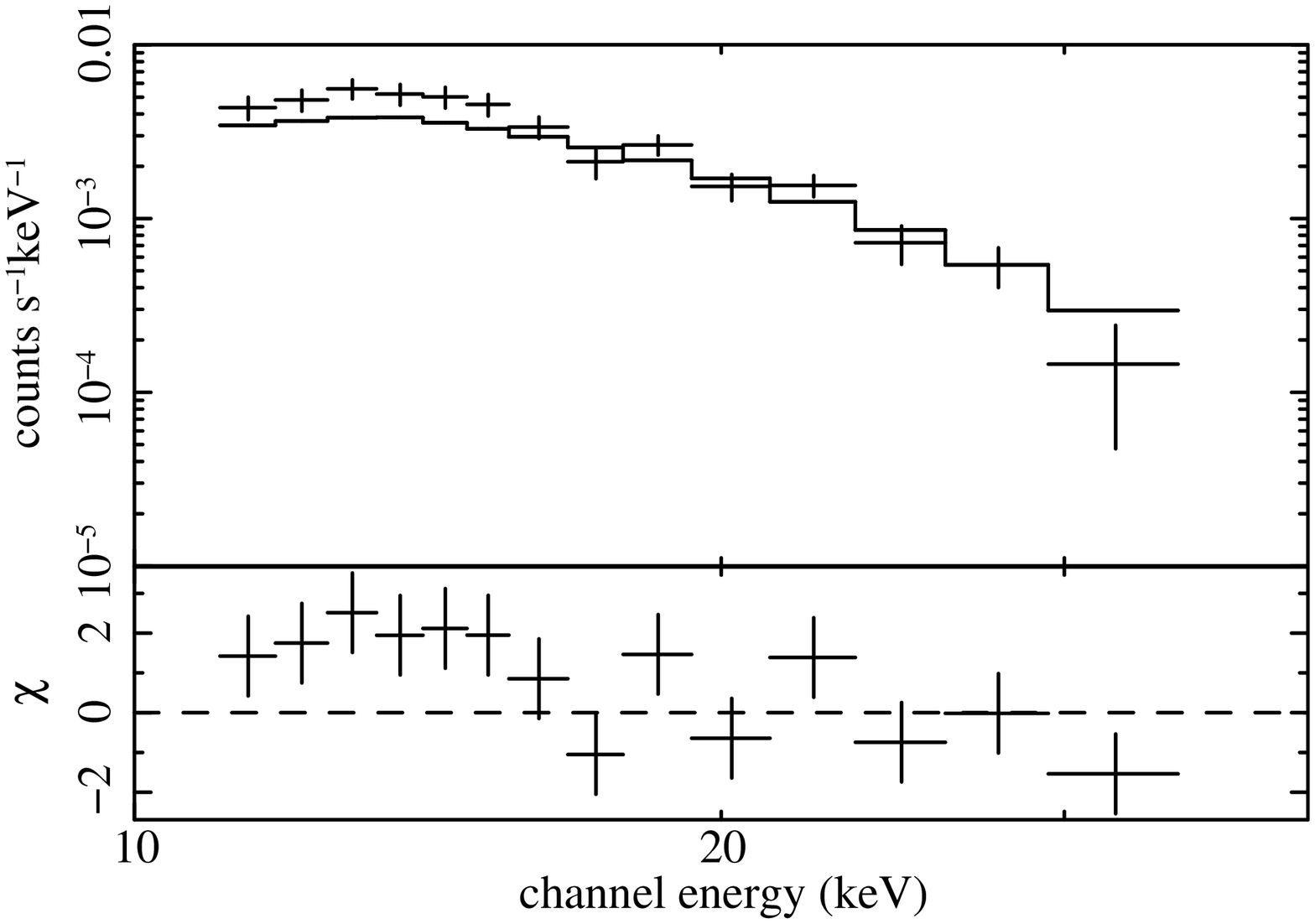}
\FigureFile(80mm,40mm){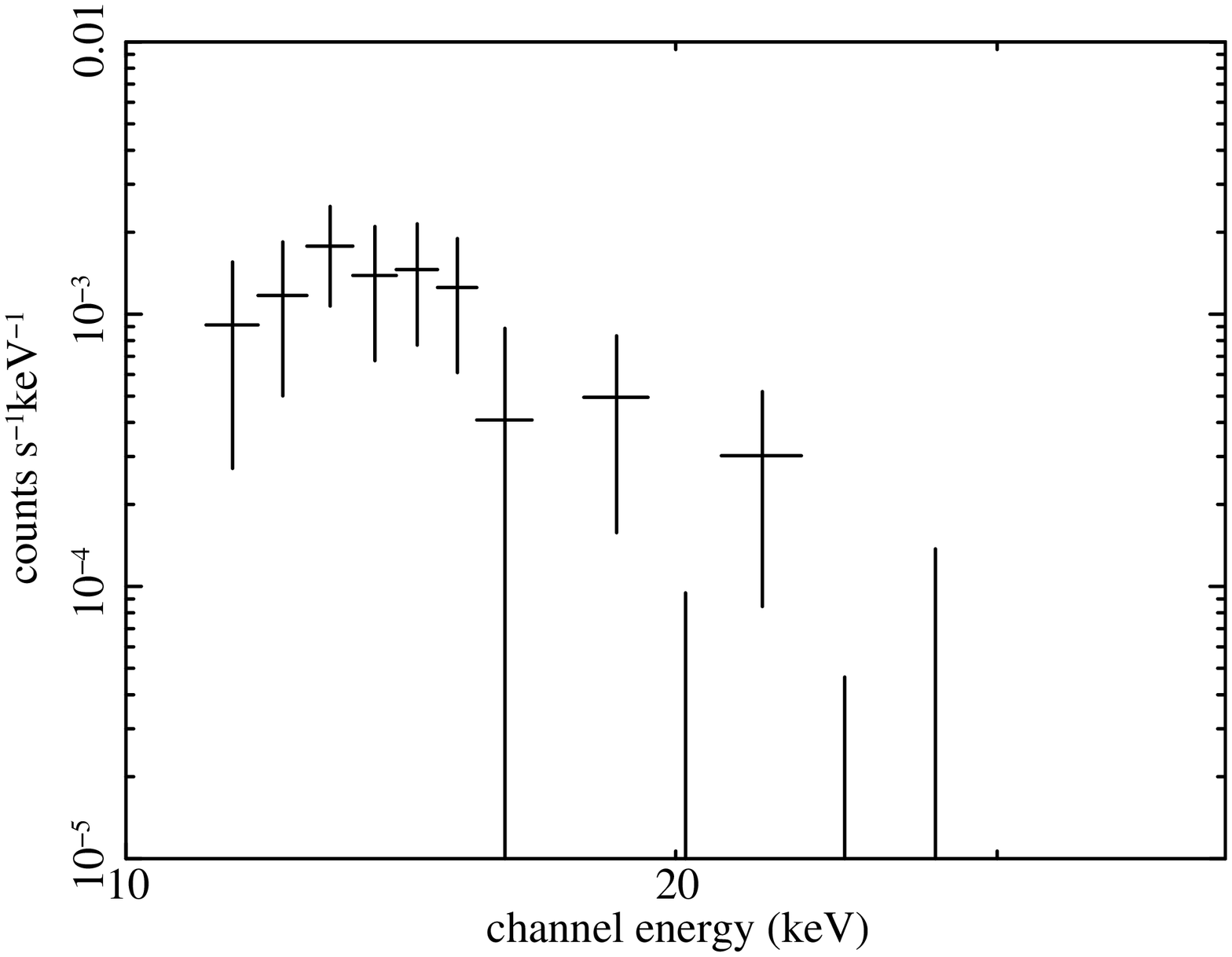}
\caption{Top: Background-subtracted PIN spectra for background (left)
and rim (right) observations.
Solid lines represent
the best-fit CXB component.
Bottom: Residuals of the rim data from the CXB component.
Note that the systematic errors are not included.}
\label{fig:hxd_spec}
\end{center}
\end{figure}

We first checked the background observations.
Figure~\ref{fig:hxd_spec} (top-left) is 
the NXB-subtracted spectrum.
Any residuals should be due to the cosmic X-ray background (CXB).
We thus fit the 15--40~keV spectrum with a CXB model\footnote{%
see http://heasarc.gsfc.nasa.gov/docs/suzaku/analysis/pin\_cxb.html .}
by HEAO-1 \citep{boldt1987}
which has a photon index of 1.29 and a cut-off energy of 40~keV.
We introduced a scaling factor to normalize the spectrum.
The best-fit scaling factor value is $1.39_{-0.22}^{+0.21}$
(errors indicate single parameter 90\% confidence regions).
The measured intensity fluctuations of the CXB flux
are too small to account for this scaling factor value.
($\sim 6\%$; \cite{kushino2002}).
However, recent observations below 10~keV suggest 
a $\sim$20\% higher CXB flux than that by \citet{boldt1987},
as summarized in \citet{frontera2006}.
\citet{kokubun2007} reported that
the PIN returns a $\sim$13--15\% larger normalization than the XIS
based on the most recent calibration using the Crab Nebula.
Additionally, other papers using Suzaku data claim
a similar normalization factor
(1.15 for MCG~$-$5$-$23$-$16; \cite{reeves2007},
1.10$\pm$0.05 for MCG~$-$6$-$30$-$15; \cite{miniutti2007}).
\citet{serlemitsos2007} reported that
the XIS normalization of the Crab Nebula
agrees with previous satellite results \citep{toor1974}
within 3\% error.
The scaling factor should therefor be approximately
1.2 $\times $ 1.15 = 1.38,
which is consistent with our best-fit value.
We fixed the CXB component at
the value by \citet{boldt1987} times 1.39,
hearafter.
Observations using Ginga indicate that excess emission local to 
SN~1006 (Lupus region) is restricted to soft X-ray band \citep{ozaki1994},
consistent with our results.

Figure~\ref{fig:hxd_spec} (top-right) shows the NXB-subtracted spectrum
from the rim observations.
We applied the CXB component
determined from the background observations
(see the previous paragraph).
In the lower energy band, 
there seem to be some positive residuals in the 10--15~keV band,
as shown in the bottom panel of figure~\ref{fig:hxd_spec},
whereas we detected no signals from other energy bands.
Note that systematic errors are not included in these figures.
We added power-law component in order to
estimate the residual flux.
We fixed the photon index of the power-law component to be 3.0
and obtained a flux of 2.7$_{-1.0}^{+0.8}\times 10^{-5}$
ph~cm$^{-2}$s$^{-1}$keV$^{-1}$ in 10--15~keV
when no systematic errors are taken into account.
We then checked the influence of systematic errors
using the following process.
Based on the HXD team report that the 90\% systematic uncertainty
of the background model in 15-40 keV is about 4\%
(figure 2 in appendix of Suzaku-Memo 2006-42), 
we constructed NXB spectra with 4\% higher count rates.
As before, we fit the NXB-subtracted spectrum with a CXB model.
A significant signal is no longer detected in any energy band.
The detection significance in the 10--15~keV band drops to 1.2$\sigma$.
The resulting 90\% upper limits in several energy bands are listed in 
table~\ref{tab:hxd_para}.
The best-fit flux for the 10--15~keV spectrum was 
2.7$_{-1.0}^{+0.8}$ (statistics) $\pm 3.6$ (systematic)
$\times 10^{-5}$ ph~cm$^{-2}$s$^{-1}$keV$^{-1}$.

\subsection{Wide band spectra}
\label{sec:wide-band}

\begin{figure}
\begin{center}
\FigureFile(100mm,40mm){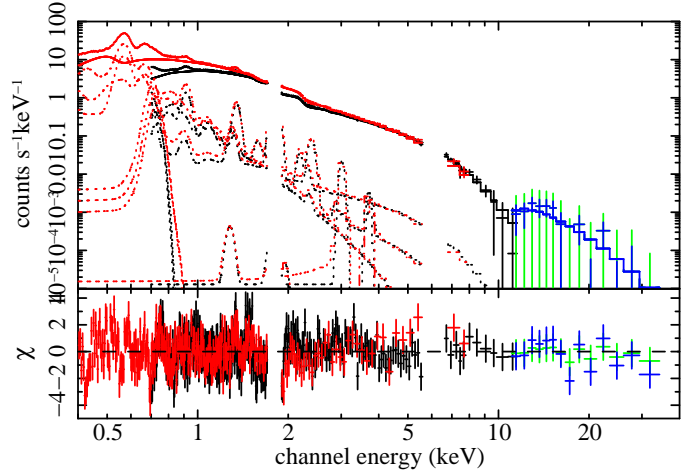}
\caption{Background-subtracted
wide band spectra of XIS (black and red) and PIN
(blue; without systematic errors, green; with systematic errors).
The solid and dotted lines are the best-fit model for the XIS spectra
(see table~\ref{tab:xis_para}).
The CXB component is also subtracted from the PIN data,
thus the PIN data are basically same as
the bottom panel of figure~\ref{fig:hxd_spec}.}
\label{fig:wideband}
\end{center}
\end{figure}

Figure~\ref{fig:wideband} shows the wide-band spectra
in the 0.4--40.0~keV band.
The CXB component for the PIN data described in \S\ref{sec:hxd-spec}
has already been subtracted.
The solid and dotted lines represent the best-fit model
(thermal + {\it srcut} model) in table~\ref{tab:xis_para}.
The normalization constant between XIS and PIN has been fixed at 1.15
\citep{kokubun2007}.
The nonthermal component is smoothly connected from the XIS band
to the PIN band,
although the systematic error of PIN prevented detection.
A more quantitative evaluation of the spectrum above 10 keV
might be possible 
with an improved NXB model of HXD PIN.

\section{Discussion}
\label{sec:discuss}

We detected both thermal and nonthermal emission
from the rims of SN~1006.
The nonthermal component is described substantially better
using an {\it srcut} model 
than a power-law model.
The roll-off energy is $5.69 (5.67$--$5.71) \times 10^{16}$~Hz
($\cong 0.23$~keV),
the most precise measurement to date.
This value is very close to that found for Tycho's remnant
($\sim 0.3$~keV; \cite{cassamchenai2007}).
The upper limit on the flux above 10 keV with HXD PIN is more stringent
 than previous observations \citep{kalemci2006}.
The roll-off energy is slightly lower than previous results
for the rim ($2.6_{-0.7}^{+0.7}\times 10^{17}$~Hz; \cite{bamba2003}),
and larger than those of inner regions
($\sim 2\times 10^{16}$~Hz; \cite{rothenflug2004}),
but consistent with our use of an integrated spectrum of the SNR.
Including high temperature components also 
would reduce roll-off energy.
We treated the flux density at 1~GHz as a free parameter,
and the result (16~Jy) is roughly consistent with the radio result
($\sim$19~Jy; \cite{green2006}).

The {\it srcut} model have only three parameters,
the flux and spectral index at 1~GHz, and the roll-off frequency.
The former two parameters are well determined by radio observations.
The systematics are then governed
only by the small uncertainties in these two
parameters measured in the radio,
and more significantly by the question of applicability of the model.
The {\it srcut} model is a simple approximation
of the emission from accelerated electrons.
Recently, some papers report the acceleration is so efficient
that the spectral shape deviates from the {\it srcut} model
(e.g., \cite{cassamchenai2007}).
More precise model for the emission from the accelerated electrons
is needed.
The fixed parameters in thermal plasma models also
make systematic uncertainty.

The magnetic field strength can be estimated 
using the upper limit in the TeV band
($3.5\times 10^{-12}$~ergs~cm$^{-2}$s$^{-1}$ in the 0.26--10~TeV;
\cite{aharonian2005,aharonian1997})
under the synchrotron/inverse Compton
scenario (in which the TeV photons are produced via the
scattering of cosmic microwave background photons by the TeV electrons).
The estimated lower limit of the magnetic field is
$\sim 25~\mu$G,
which is consistent with those estimated in other ways
(40~$\mu$G; \cite{allen2001,bamba2003}).

The roll-off energy $\nu_{roll}$ is described as
\begin{equation}
\nu_{roll} = 1.6\times 10^{16} \left(\frac{B}{10~\mu {\rm G}}\right)
\left(\frac{E_e}{10~{\rm TeV}}\right)^2\ \ ,
\label{eq:roll}
\end{equation}
where $E_e$ and $B$ represent
the maximum energy of accelerated electrons and the magnetic field strength
respectively
\citep{reynolds1998,reynolds1999}.
The maximum energy of electrons is estimated to be
9.4~TeV,
under the assumption that
the downstream magnetic field strength is 40~$\mu$G,
The time scale of synchrotron loss $\tau_{loss}$ is 
\begin{equation}
\tau_{loss} = 2.0\times 10^4 \left(\frac{B}{10~\mu{\rm G}}\right)^{-2}
\left(\frac{E_e}{10~{\rm TeV}}\right)^{-1}\ \ {\rm [yrs]}\ \ ,
\end{equation}
then $\sim$1300~yrs in this case,
roughly same as the SNR age.
This result implies that
the acceleration system in SN~1006 is changing
from age-limited acceleration to synchrotron-loss limited acceleration.

The roll-off energy is significantly higher in the NE region,
which is consistent with the XMM-Newton result \citep{rothenflug2004}.
This result implies that the magnetic field is stronger 
and/or electrons are accelerated to higher energy there.
In other words,
the acceleration efficiency is higher in the NE region.
Assuming a magnetic field strength of 40~$\mu$G again,
we derived the maximum energy of electrons with eq.(\ref{eq:roll})
to be 10~TeV (NE) and 8.5~TeV (SW),
respectively.

\section{Summary}
\label{sec:summary}

We have presented
wide band spectra of SN~1006
taking advantage of the excellent statistics and the low background of the Suzaku
XIS and HXD.
The nonthermal component has a roll-off energy of
$5.7\times 10^{16}$~Hz.
The HXD PIN provides the most stringent upper limit above 10~keV.
The NE rim has the higher roll-off energy than the SW rim.

\section*{Acknowledgements}

We thank W. Hofmann
for useful comments.
We appreciate all members of the Suzaku team,
with our particular thanks to the HXD hardware team members.
The authors also thank K.~Makishima, M.~Ishida, and Y.~Maeda
for their fruitful comments.
A.~Bamba, H.~Nakajima, H.~Takahashi, T.~Tanaka, and H.~Yamaguchi
are supported by JSPS Research Fellowship for Young Scientists.

\appendix

\section{Selection of HXD PIN background models}
\label{sec:appendix}

Two types of background models are provided by HXD team,
bgd-a and bgd-d,
and a comparison of these two models is useful
for validating the background model for observations of interest.
The NXB model of bgd-a was constructed by sorting
earth occultation data using two parameters, PIN upper discriminator (PINUD)
count rate and PINUD build-up counts.
The former is used to take the COR dependence into account.
The latter is a convolution of the PINUD counts 
with an exponential decay function with a certain time constant, and thus
takes into account the activation-induced background component 
gradually decaying after
the passage of South Atlantic Anomaly
(Watanabe et al.\ Suzaku-Memo 2007-01).
The NXB model of bgd-d was constructed so as to reproduce the light
curve of the earth occultation data with an empirical model.
The model parameters are such as PINUD, PINUD build up with 4 time
constants, GSO high energy counts, and the angle between the HXD field
of view and the magnetic field.
The latter three parameters are used to model the activation-induced
background variation.
A set of monthly model parameters are determined
by fitting the light curve of the Earth occultation data from each
month (Fukazawa et al.\ Suzaku-Memo 2007-02).
Basic features of reproducibility of both models are reported 
in Mizuno et al. (Suzaku-Memo 2006-42).

\begin{figure}
\begin{center}
\FigureFile(80mm,40mm){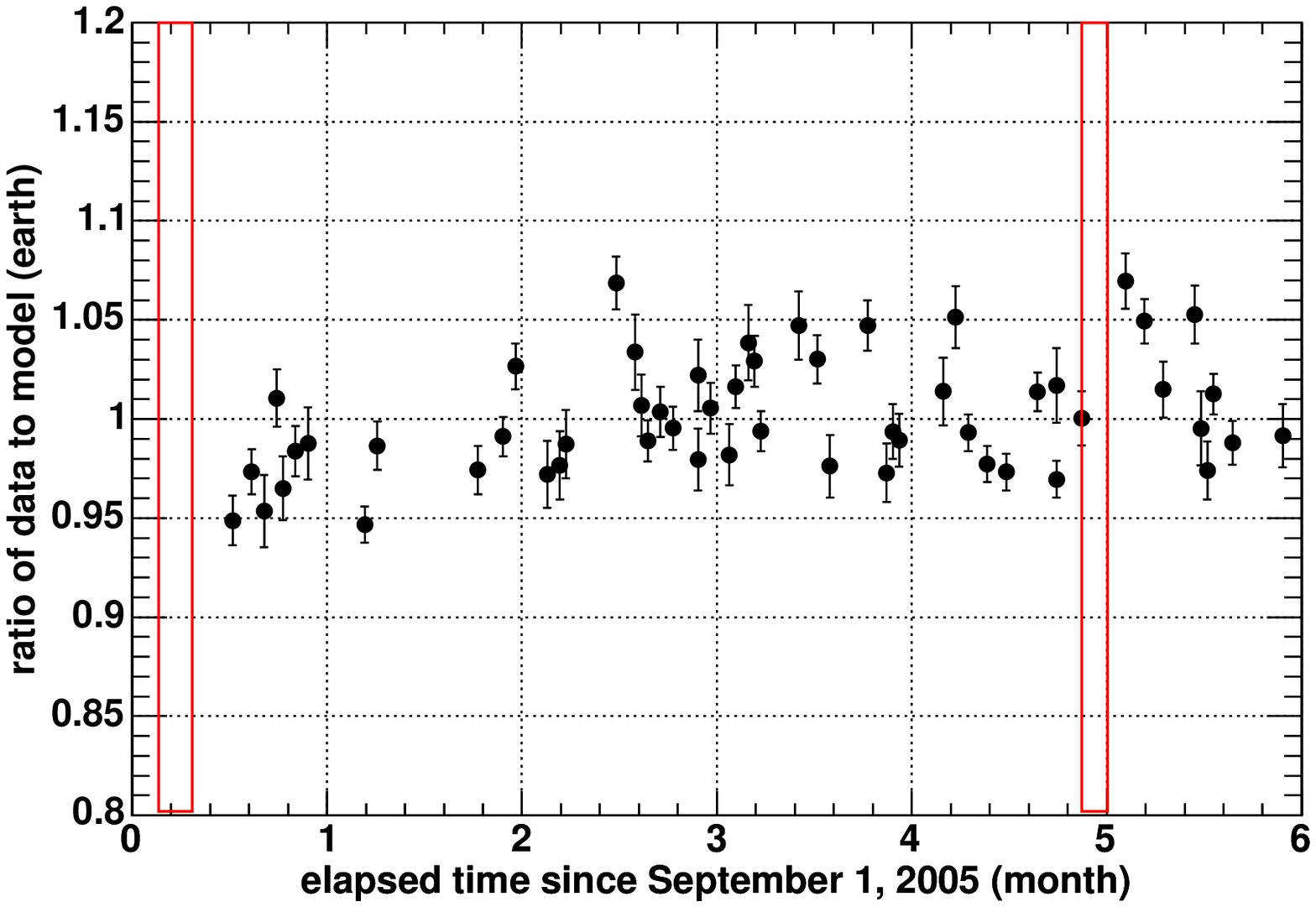}
\FigureFile(80mm,40mm){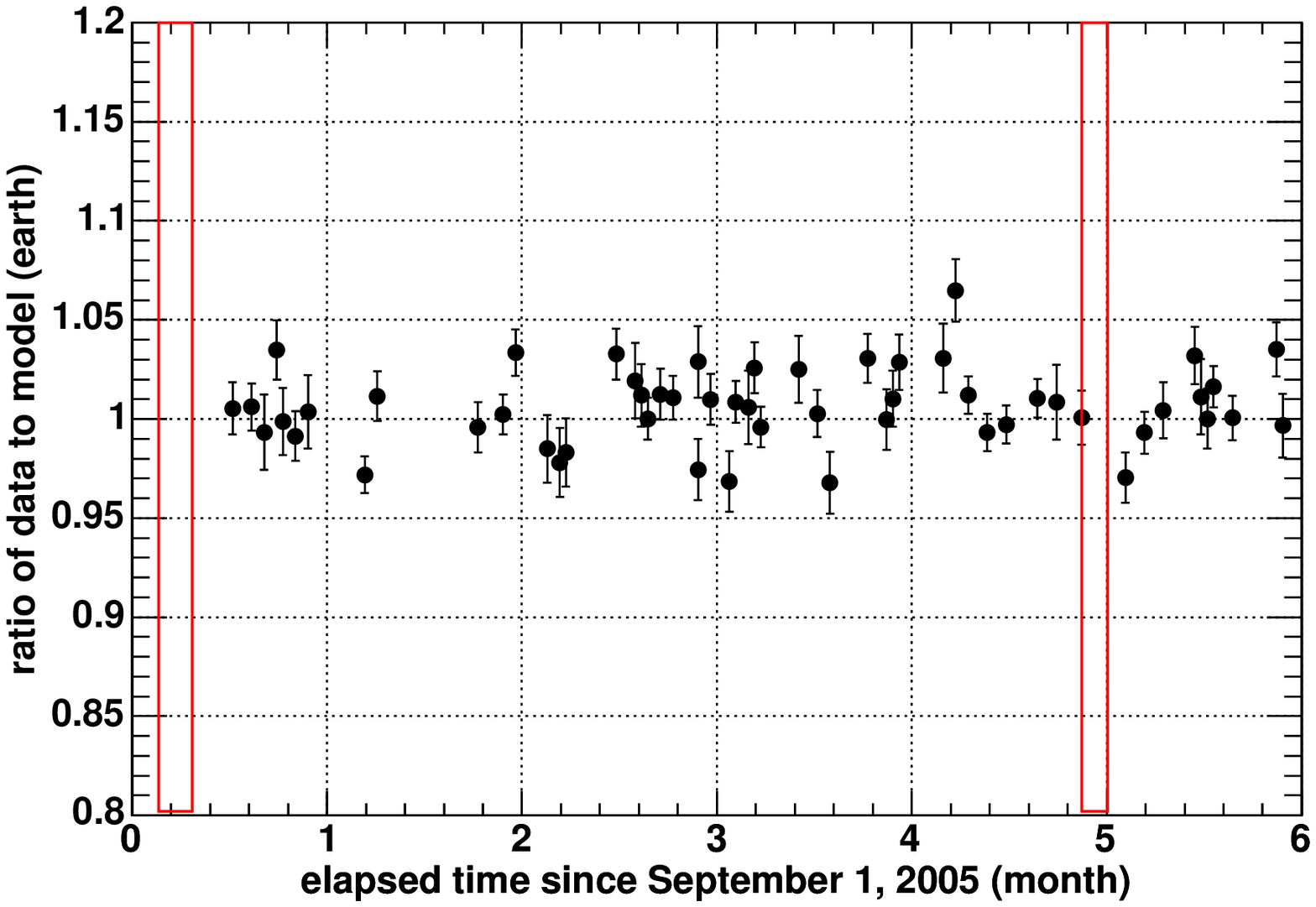}
\caption{The ratio between the 15--40~keV count rate of
data and bgd-a (left) and bgd-d (right) during the earth occultation,
from 2005 September to 2006 March.
Red boxes represent the observation periods for SN~1006.}
\label{fig:earth}
\end{center}
\end{figure}

We found that,
for observations of SN1006 in January 2006,
bgd-a systematically gives
$\sim$7\% lower background flux than bgd-d (or vise versa).
Although the reproducibility of the background model can be checked
by looking at the earth occultation data in principle,
observations of SN1006 unfortunately lacked the
exposure to the earth.
We therefore compared the 15--40~keV count rates
of all the earth occultation data
and background model prediction (both bgd-a and bgd-d) 
with exposure more than 10~ks from 2005 September to 2006 March.
Figure~\ref{fig:earth} shows the ratio
between data and the NXB model prediction,
which should be 1.
We note that both bgd-a and bgd-d reproduce the data well within 
$\sim$5\%, except for the period from
the end of 2006 January to the beginning of 2006 February
(5th month from 2005 September),
where bgd-a gives a ratio larger than 1 by about 5\%.
In the figure, the period of SN1006 observations is indicated by red rectangles
and 4 out of 8 observations are in the latter region.
Since these 4 observations might have suffered from this anomaly,
we adopted bgd-d as the NXB model in our analysis.

\begin{figure}
\begin{center}
\FigureFile(80mm,50mm){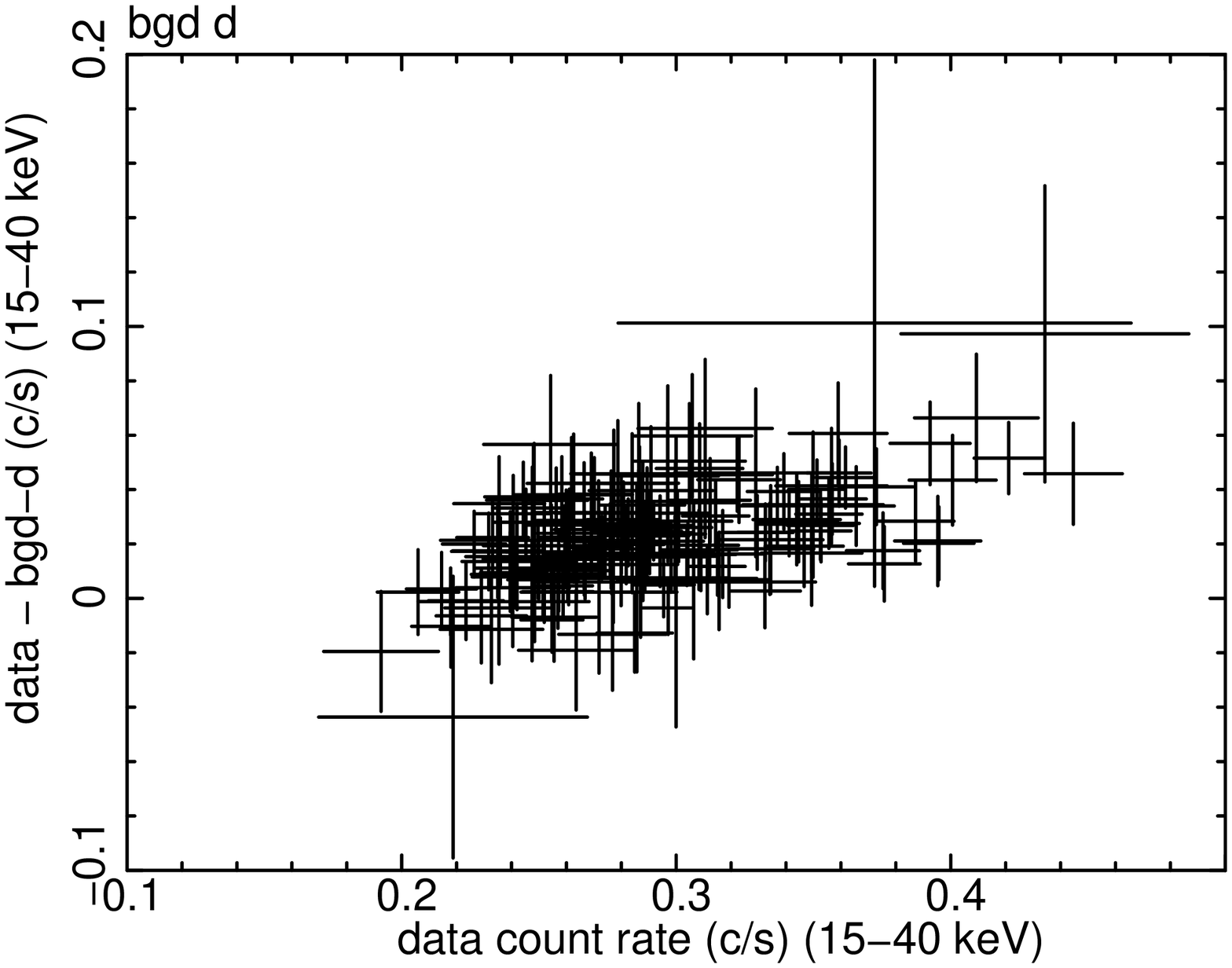}
\FigureFile(60mm,50mm){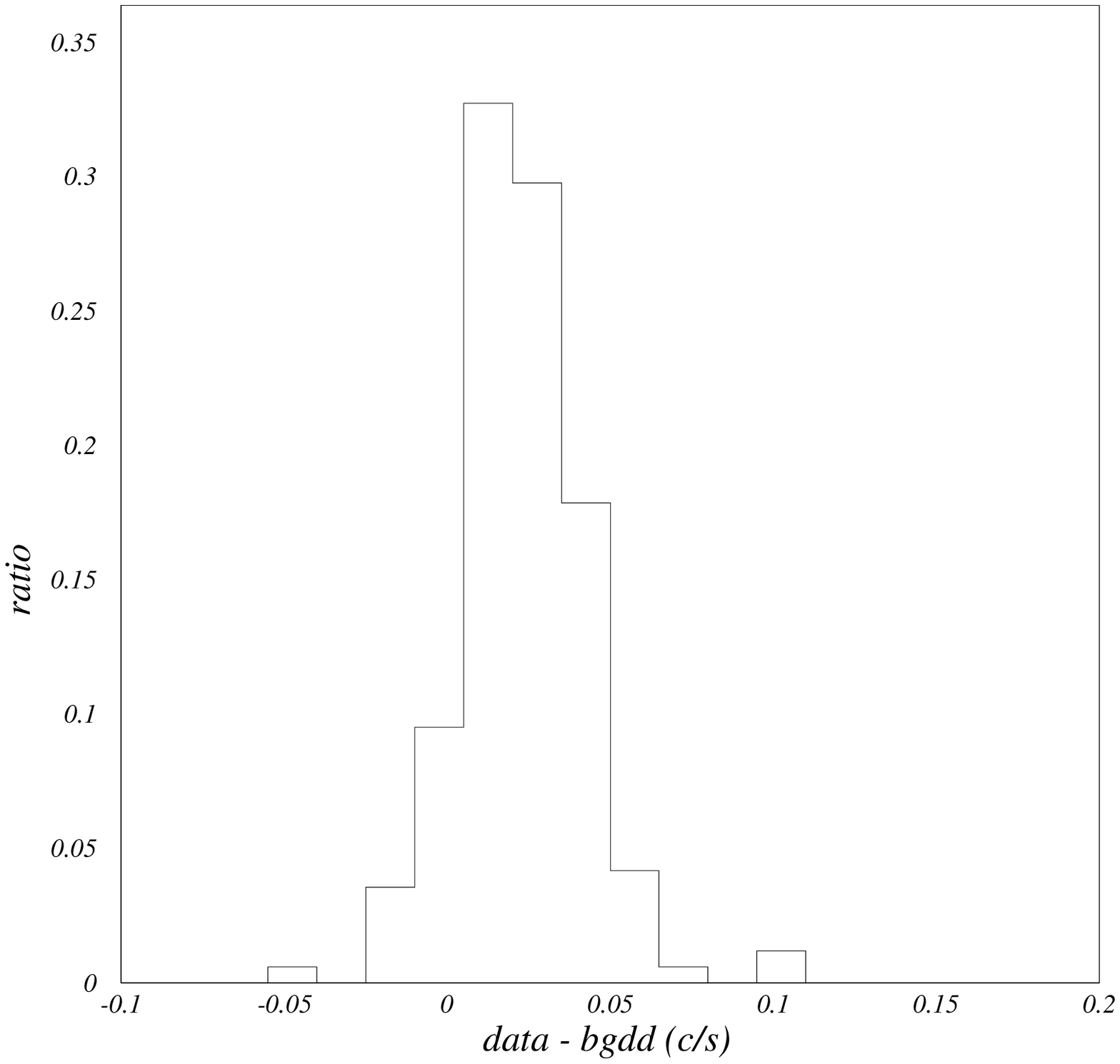}
\caption{(left) PIN data count rate vs. data$-$bgd-d model count rate.
The energy band is 15--40~keV.
(right) Histogram of data$-$bgd count rates with the bgd-d model.}
\label{fig:pin_counts}
\end{center}
\end{figure}

As the next step,
we checked the background reproducibility in short periods
by comparing the
data count rate vs. data$-$bgd-d count rate for each 4096~sec interval
in our 8 observations,
which should be independent of the data count rate.
The left panel of figure~\ref{fig:pin_counts}
shows the count rate vs. data $-$ bgd-d in the 15--40~keV band.
The data have a slightly positive correlation with the bgd-d model,
but it is moderate.
The right panel of figure~\ref{fig:pin_counts} shows that
the data$-$bgd-d hava a small dispersion
at the level of $\sim 2\times 10^{-2}$~count~s$^{-1}$.
This value is roughly similar to that of the CXB emission.
Therefore, we concluded that
the bgd-d model reproduces the real NXB in our observations
and used it in this paper.

Note that the large residual found here is not generally seen 
and bgd-a and bgd-d provide us with consistent result most of the times,
as shown by \citet{reeves2007},
\citet{miniutti2007},
and \citet{kataoka2007}.

\onecolumn

\begin{table}[hbtp]
\begin{center}
\caption{Observation Log.}
\label{tab:obslog}
\begin{tabular}{p{5pc}cccc}\hline\hline
 & ObsID & Date & Aim point & Exposure (XIS/HXD) \\
 & & (yyyy/mm/dd) & (RA, Dec.) & (ks) \\
\hline
NE\_bg\dotfill & 100019010 & 2005/09/04 & (226.7005, $-$41.4006) & 50/31 \\
NE\dotfill & 100019020 & 2005/09/09 & (225.9608, $-$41.7805) & 43/31 \\
SW1\dotfill & 100019030 & 2005/09/10 & (225.5010, $-$42.0706) & 31/22 \\
SW\_bg1\dotfill & 100019040 & 2005/09/11 & (224.6505, $-$42.4005) & 27/21 \\
SW2\dotfill & 100019050 & 2006/01/26 & (225.4998, $-$42.0701) & 31/21 \\
SW\_bg2\dotfill & 100019060 & 2006/01/26 & (224.6496, $-$42.4009) & 30/25 \\
SE\dotfill & 500016010 & 2006/01/30 & (225.8686, $-$42.0508) & ---$^{a}$/48 \\
NW\dotfill & 500017010 & 2006/01/31 & (225.6397, $-$41.7993) & ---$^{a}$/51 \\
\hline
\multicolumn{4}{l}{$^{a}$: Data is not used in our analysis.}\\
\end{tabular}
\end{center}
\end{table}

\begin{table}
\begin{center}
\caption{XIS spectral fitting parameters$^{a}$.}
\label{tab:xis_para}
\begin{tabular}{p{20pc}cc}\hline\hline
Parameters & thermal$^b$ + power-law & thermal$^b$ + {\it srcut} \\
\hline
$N_{\rm H}$ [cm$^{-2}$]\dotfill & \multicolumn{2}{c}{$6.8\times 10^{20}$ (fixed) } \\
VNEI~1 (ejecta 1)\\
\hspace*{3mm} $kT$ [keV]\dotfill & \multicolumn{2}{c}{1.2 (fixed$^b$)} \\
\hspace*{3mm} $n_{\rm O}n_{\rm e}V$ [cm$^{-3}$]\dotfill & 2.86 (2.45--3.06) $\times 10^{52}$ & 4.19 (4.05--4.32) $\times 10^{52}$ \\
VNEI~2 (ejecta 2)\\
\hspace*{3mm} $kT$ [keV]\dotfill & \multicolumn{2}{c}{1.9 (fixed$^b$)} \\
\hspace*{3mm} [S/O]\dotfill & \multicolumn{2}{c}{2.7 (fixed)} \\
\hspace*{3mm} $n_{\rm O}n_{\rm e}V$ [cm$^{-3}$]\dotfill & 8.43 (8.18--8.55) $\times 10^{53}$ & 3.82 (3.77--3.89) $\times 10^{53}$ \\
NEI (ISM)\\
\hspace*{3mm} $kT$ [keV]\dotfill & \multicolumn{2}{c}{0.45 (fixed)} \\
\hspace*{3mm} $n_e$t [cm$^{-3}$s]\dotfill & \multicolumn{2}{c}{5.7$\times 10^9$ (fixed)} \\
\hspace*{3mm} $n_{\rm H}n_{\rm e}V$ [cm$^{-3}$]\dotfill & 1.14 (1.06--1.21) $\times 10^{56}$ & 3.45 (3.43--3.48) $\times 10^{56}$ \\
Nonthermal component\\
\hspace*{3mm} $\Gamma$/$\nu_{roll}$ [---/Hz]\dotfill & 2.73 (2.72--2.74) & 5.69 (5.67--5.71) $\times 10^{16}$ \\
\hspace*{3mm}Norm [ph~keV$^{-1}$cm$^{-2}$s$^{-1}$ at 1~keV/Jy at 1~GHz]\dotfill & 4.05 (4.04--4.07) $\times 10^{-2}$ & 16.2 (16.1--16.3) \\
Gain offset for FI [eV]\dotfill & 3.9 & $-$1.4 \\
Gain offset for BI [eV]\dotfill & $-$5.0 & $-$4.0 \\
$\chi^2$/d.o.f.\dotfill & 2200/588 & 857/588 \\
\hline
\multicolumn{3}{l}
{$^a$: Parentheses indicate single parameter 90\% confidence regions.}\\
\multicolumn{3}{l}
{$^b$: Thermal parameters are fixed following \citet{yamaguchi2007}.}\\
\end{tabular}
\end{center}
\end{table}

\begin{table}
\begin{center}
\caption{XIS spectral fitting parameters for NE and SW regions$^a$.}
\label{tab:xis_para_parts}
\begin{tabular}{p{12pc}cc}\hline\hline
Parameters & NE & SW \\
\hline
VNEI~1 (ejecta 1)\\
\hspace*{3mm} $n_{\rm O}n_{\rm e}V$ [cm$^{-3}$]\dotfill & 1.20 (1.15--1.39) $\times 10^{52}$ & 2.76 (2.61--2.83) $\times 10^{52}$ \\
VNEI~2 (ejecta 2)\\
\hspace*{3mm} $n_{\rm O}n_{\rm e}V$ [cm$^{-3}$]\dotfill & 2.10 (2.06--2.15) $\times 10^{53}$ & 1.88 (1.76--1.95) $\times 10^{53}$ \\
NEI~1 (ISM)\\
\hspace*{3mm} $n_{\rm H}n_{\rm e}V$ [cm$^{-3}$]\dotfill & 1.09 (1.08--1.13) $\times 10^{56}$ & 2.19 (2.16--2.24) $\times 10^{56}$ \\
{\it srcut}\\
\hspace*{3mm} $\nu_{roll}$ [Hz]\dotfill & 6.66 (6.58--6.69) $\times 10^{16}$ & 4.68 (4.64--4.73) $\times 10^{16}$ \\
\hspace*{3mm}Norm [Jy at 1~GHz]\dotfill & 7.72 (7.69--7.74) & 8.33 (8.03--8.57) \\
Gain offset for FI [eV]\dotfill & $-$2.0 & $-$2.9 \\
Gain offset for BI [eV]\dotfill &  $-$6.0 & $-$4.0 \\
$\chi^2$/d.o.f.\dotfill & 553/338 & 527/368 \\
\hline
\multicolumn{3}{l}
{$^a$: Parentheses indicate single parameter 90\% confidence regions.}\\
\end{tabular}
\end{center}
\end{table}

\begin{table}
\begin{center}
\caption{90\% upper-limits of nonthermal emission}
\label{tab:hxd_para}
\begin{tabular}{p{8pc}c}\hline\hline
Energy band & 90\% upper-limit \\
 & [ph~cm$^{-2}$s$^{-1}$keV$^{-1}$] \\
\hline
10 -- 15 keV \dotfill & $7.1\times 10^{-5}$ \\
15 -- 20 keV \dotfill & $2.3\times 10^{-5}$ \\
20 -- 40 keV \dotfill & $8.8\times 10^{-6}$ \\
\hline
\end{tabular}
\end{center}
\end{table}

\end{document}